\documentclass[twocolumn,showpacs,preprintnumbers,amsmath,amssymb]{revtex4}

\usepackage{graphicx}
\usepackage{dcolumn}
\usepackage{bm}

\begin{document}


\title{Critical Space-Time Networks and Geometric Phase Transitions \\ from Frustrated Edge Antiferromagnetism}


\author{Carlo A. Trugenberger}
\email{ca.trugenberger@bluewin.ch}
\affiliation{%
SwissScientific, chemin Diodati 10, CH-1223 Cologny, Switzerland
}%

\date{\today}

\begin{abstract}
Recently I proposed a simple dynamical network model for discrete space-time which self-organizes as a graph with Hausdorff dimension $d_H=4$. The model has a geometric quantum phase transition with disorder parameter $(d_H-d_s)$ where $d_s$ is the spectral dimension of the dynamical graph. Self-organization in this network model is based on a competition between a ferromagnetic Ising model for vertices and an antiferromagnetic Ising model for edges. In this paper I solve a toy version of this model defined on a bipartite graph in the mean field approximation. I show that the geometric phase transition corresponds exactly to the antiferromagnetic transition for edges, the dimensional disorder parameter of the former being mapped to the staggered magnetization order parameter of the latter. The model has a critical point with long-range correlations between edges, where a continuum random geometry can be defined, exactly as in Kazakov's famed 2D random lattice Ising model but now in any number of dimensions. 
\end{abstract}
\pacs{04.60.-m; 89.75-Hc}
\maketitle

\section{Introduction}
Networks play a role in several of the major contenders for a theory of quantum geometries. In the spin foam approach \cite{spinfoam} to loop quantum gravity \cite{loop} spin networks represent the quantum states of the gravity on a given manifold. In the causal dynamical triangulation approach (CDT) \cite{triang}, triangulations of space-time play the role of a lattice regularization of the Einstein-Hilbert action used to search for a non-trivial ultraviolet (UV) renormalization group fixed point defining quantum gravity in a non-perturbative sense, an approach that goes under the name of asymptotic safety \cite{safety}. Causal sets \cite{sorkin} and energetic causal sets \cite{smolin} are attempt to build geometries from networks of causal relations. Quantum graphity models \cite{graphity} attempt the same starting from graphs. Finally, a very recent approach \cite{bianconi} is the idea of considering discrete space-time as a quantum network, obtained by growing simplicial complexes. 

In a recent publication \cite{trug} I proposed a different approach to quantum gravity in which networks play an even greater role. The idea is to turn things around and, instead of starting from the correct infrared (IR) geometric variables and use discrete space-time just as a regularization artefact, as in the CDT approach, to posit purely combinatoric information bits as the fundamental UV quantum variables of gravity and have these self-organize so that geometry and general relativity emerge.  

Of course this presupposes the existence of a model having an UV fixed point corresponding to disordered bits and an IR limit in which vertices and edges self-organize to form a network with the topology of discrete space-time. In \cite{trug} I have proposed exactly such a model, suitably generalizing Kazakov's famed random lattice Ising model \cite{kazakov}. As I have shown in \cite{trug} there is ample numerical and analytical evidence 
supporting the fact that the Hausdorff dimension of the emergent space-time in this model is {\it predicted} to be 4, the upper critical dimension of the Ising model \cite{zin}. 

This model distinguishes itself from previous approaches in two major respects. First of all, it is much more radical, in the sense that no constraints at all, like the requirements of triangulations, causal sprinklings of Lorentz manifolds or simplicial complexes, are imposed on the discrete basic structure: this is completely determined by self-organization driven by a minimum energy principle alone. Secondly, nothing is assumed a priori about time: contrary to most previous authors I believe time should not be assumed by imposing causality from the very beginning, but should be, rather, "explained" by the model itself. Admittedly, though, I am still quite short of this goal. The idea, however, is to explore if space-time can emerge from the simplest combinatoric variables with no other driving force than the competition arising from frustrated Ising Hamiltonians. 

\section{The model}
The model is formulated in terms of $N$ spin 1/2 bits $s_i = \pm 1$, $i=1\dots N$, whose values indicate the presence/absence of (Euclidean) space-time, respectively and $N(N-1)/2$ symmetric spin 1/2 bits $w_{ij}=w_{ji}=0,1$, $i,j=1\dots N$, whose values denote a connection (1) between space-time vertices $s_i$ and $s_j$ or the absence thereof (0) ($w_{ii}=0$)
\begin{equation}
{\cal H} = {J\over 2} \ \sum_{i\ne j} \sum_{k\ne i \atop k\ne j} w_{ik}w_{kj} - {1\over 2} \sum_{i\ne j} s_i w_{ij} s_j \ , 
\label{one}
\end{equation}
where $J$ is the unique dimensionless coupling. I use units in which $\hbar=1$, $c=1$ and all energies are measured in units of the standard Ising coupling, second term in (\ref{one}), which is set to one for simplicity of presentation. 

The second term in this energy function is the standard ferromagnetic Ising model. If the links $w_{ij}$ would be uniformly drawn from random adjacency matrices of degree 4, the model would be exactly Kazakov's random lattice Ising model in two dimensions \cite{kazakov}. The first term in the energy function, on the other side, is simply a nearest-neighbours (sharing a common vertex) antiferromagnetic Ising model for the link spins. The generalizations with respect to Kazakov's model, thus consist in dropping the restriction to degree 4 and drawing the random adjacency matrices from a Gaussian distribution. 

The driver of self-organization in this model is the competition between the vertex ferromagnetic coupling and the link antiferromagnetic one, creating "link frustration". Indeed the vertex ferromagnetic coupling favours the creation of many links (in a vertex aligned configuration) while the antiferromagnetic link coupling tends to suppress such links. As I have shown in \cite{trug}, for $J=1/(4d-1)$, with integer $d$, the compromise is a $2d$-regular ground state graph with power-law extension, exactly what one would expect for a discretized space-time. The spectral and Hausdorff dimensions \cite{dim} of this ``space-time graph" are determined by the unique dimensionless coupling $J$ of the model. The spectral dimension is simply $d$ while all evidence supports the fixed value 4 for the Hausdorff dimension $d_H$ in the whole range of couplings between the two quantum phase transitions at $d=1$, corresponding to the lower critical dimension of the Ising model, where space-time vertices themselves become disordered and $d=4$, the upper critical dimension of the Ising model where spectral and Hausdorff dimension start to coincide. 

\section{A toy version of the model on bipartite graphs}
The purpose of the present paper is to solve a simplified version of (\ref{one}) in the mean field approximation to elucidate the nature of the upper phase transition with disorder parameter ($d_H-d$) and to show how critical graphs with long-range correlations and power-law extension emerge as a result of frustrated edge antiferromagnetism. 

Near the upper transition all space-time vertices are aligned. I will thus assume a configuration with $s_i=+1$, $\forall i$. In this case the model (\ref{one}) reduces to a dynamical graph problem with an edge Hamiltonian
\begin{equation}
{\cal H}= {J\over 2} \sum_i \sum_{e(i)\ne f(i)} \sigma_{e(i)} \sigma_{f(i)} - {1\over 2} \sum_i \sum_{e(i)} \sigma_{e(i)} \ ,
\label{two}
\end{equation}
where $e(i)=0,1$ and $f(i)=0,1$ denote edges emanating from vertex $i$.  In this case the frustration reduces to the competition between the edge antiferromagnetic interaction (first term) and an external edge ``magnetic field" generated by the aligned space-time vertices (second term). In this space-time-aligned phase, the model can be thought of as being defined on the complete graph on $N$ vertices, the (+1) edges defining a dynamically generated subgraph representing the emergent space-time. The connectivity of this subgraph is dynamically determined by the coupling constant $J$: for $J=1/(4d-1)$ with integer $d$ the subgraph is $2d$-regular. The toy model I will consider in this paper is defined by the same Hamiltonian (\ref{two}) but restricted on a connected, bipartite $2d$-regular graph, while $J$ is left free. As I will now show, this model is solvable in the mean field approximation. 

As a first step I will introduce a more familiar notation by defining standard spins $\xi_{e(i)} = \pm 1$ as $\xi_{e(i)}= 2\sigma_{e(i)}-1$. The Hamiltonian (\ref{two}) reduces then to 
\begin{eqnarray}
{\cal H} &&= {J\over 8} \sum_i \sum_{e(i)\ne f(i)} \xi_{e(i)} \xi_{f(i)} - h \sum_i \sum_{e(i)} \xi_{e(i)} \ ,
\nonumber \\
h &&= {1-(2d-1) J \over 4}  \ .
\label{three}
\end{eqnarray}

The second step is to invoke K\"onig's edge colouring theorem \cite{konig}, which states that the edge chromatic number of any bipartite graph equals its maximum vertex degree, in this case $2d$ since the graph is assumed $2d$-regular. This means that at every vertex one can colour the incident edges with exactly $2d$ different colours, without two edges of the same colour ever touching. Let me now further subdivide the $2d$ colours into two types of light and dark colours: at every vertex there will be exactly $d$ light colours and $d$ dark colours. Exactly like one standardly treats antiferromagnets on bipartite lattices by introducing two different magnetizations for the two sub-lattices I will consider configurations with two different magnetizations $m_l$ and $m_d$ (defined as usual between -1 and 1) for the two types of edge colours and define 
\begin{equation}
m = {m_d + m_l\over 2} \ ,\qquad \qquad m_s = {m_d - m_l\over 2} \ ,
\label{four}
\end{equation}
as the magnetization and the staggered magnetization, respectively. The original $2d$-regular graph corresponds clearly to a maximally ``ferromagnetic configuration" $m=1$, $m_s=0$. The maximally ``antiferromagnetic configuration" $m=0$, $m_s=1$, instead corresponds to a regular connected sub-graph. In between these two extremes there are intermediate configurations in which each dark-coloured edge has a probability $p_d= (1+m_d)/2$ of being present in the graph, whereas the probability for light-coloured edges is $p_l= (1+m_l)/2$. 

\section{Solution of the toy model in the mean field approximation}
In the mean field approximation I will decompose each edge spin into its mean value $m_{e(i)}$ and a fluctuation around it: $\xi_{e(i)} = m_{e(i)} + \delta \xi_{e(i)}$, with $\delta \xi_{e(i)} = \xi_{e(i)} -m_{e(i)}$ and neglect terms quadratic in the fluctuations in the Hamiltonian, so that 
\begin{eqnarray}
\beta {\cal H}_{MF} &&= {\cal J} \sum_i \sum_{e(i)\ne f(i)} m_{e(i)} \xi_{f(i)} + m_{f(i)} \xi_{e(i)}  - m_{e(i)}  m_{f(i)} 
\nonumber \\
&& - H \sum_i \sum_{e(i)} \xi_{e(i)} \ ,
\label{five}
\end{eqnarray}
where $\beta =1/T$ is the inverse temperature and 
\begin{equation}
{\cal J} = \beta {J\over 8} \ , \qquad \qquad H=\beta h \ .
\label{six}
\end{equation}

At this point one can compute exactly the free energy per edge (a $2d$-regular graph on $N$ vertices has $dN$ edges by the degree sum formula). 
\begin{equation}
f_{MF} = {8\over JdN} F_{MF} = -{1\over  dN {\cal J}}\  {\rm ln} \ \sum_{\{ e \} } e^{-\beta {\cal H}_{MF}} \ .
\label{seven}
\end{equation}
I will consider first the Ansatz of configurations with exactly $d$ edges of mean $m_d$ and $d$ edges of mean $m_l$ at every vertex. Apart from an irrelevant constant, in this case the free energy becomes
\begin{eqnarray}
f_{MF} &&= - (d-1) (m_d^2 + m_l^2) - 2 d \ m_d m_l 
\nonumber \\
&&- {1\over {\cal J}} {\rm ln} \ {\rm cosh} \left[ 2{\cal J} \left( (d-1) m_d +d m_l\right) -H \right] 
\nonumber \\
&&- {1\over {\cal J}} {\rm ln} \ {\rm cosh} \left[ 2{\cal J} \left( (d-1) m_l+d m_d\right) -H \right] \ ,
\label{eight}
\end{eqnarray} 
Note that, contrary to the usual situation in staggered antiferromagnets, where only vertex-spins of the two different sub-lattices interact, here there are both interactions between the two types of light- and dark-coloured edges and amongst each type of colour also. Using (\ref{four}) the free energy can also be easily rewritten in terms of the magnetization $m$ and the staggered magnetization $m_s$,
\begin{eqnarray}
f_{MF} &&= - (4d-2) m^2 + 2 m_s^2 
\nonumber \\
&&- {1\over {\cal J}} {\rm ln} \ {\rm cosh} \left[ 2{\cal J} \left( (2d-1) m + m_s\right) -H \right] 
\nonumber \\
&&- {1\over {\cal J}} {\rm ln} \ {\rm cosh} \left[ 2{\cal J} \left( (2d-1) m_l-m_s\right) -H \right] \ ,
\label{nine}
\end{eqnarray} 

The stationarity conditions $\partial f_{MF}/\partial m=0$ and $\partial f_{MF}/\partial m_s=0$ lead to the equations
\begin{eqnarray}
m+m_s &&= {\rm th} \left[ H-2{\cal J} \left( (2d-1) m - m_s\right) \right] \ ,
\nonumber \\
m-m_s &&= {\rm th} \left[ H-2{\cal J} \left( (2d-1) m + m_s\right) \right] \ . 
\label{ten}
\end{eqnarray}
Using the following formulas for the hyperbolic tangent 
\begin{eqnarray}
{\rm th} (x+y) &&= {{\rm th}(x) + {\rm th}(y) \over 1 + {\rm th}(x) {\rm th}(y)} \ ,
\nonumber \\
{\rm th} (x-y) &&= {{\rm th}(x) - {\rm th}(y) \over 1 - {\rm th}(x) {\rm th}(y)} \ ,
\label{eleven}
\end{eqnarray}
gives then two coupled equations for the two order parameters $m$ and $m_s$ in which one of them has no explicit dependency on the magnetic field, 
\begin{eqnarray}
{2 m_s\over 1+m_s^2-m^2} &&= {\rm th} \left[ 4 {\cal J} m_s\right] \ ,
\nonumber \\
{2 m\over 1+m^2-m_s^2} &&= {\rm th} \left[  2H-4{\cal J} (2d-1) m \right] \ .
\label{twelve}
\end{eqnarray} 
The first immediate observation about these equations is that, for $H\ne 0$ and finite ${\cal J}$, the solution of the second equation always implies $m\ne 0$. Let me now consider the first equation for the staggered magnetization. The function on the left-hand side has a slope $2/(1-m^2)$ at the origin, has two extrema at $\pm \sqrt{1-m^2}$ and vanishes at $\pm \infty$. The hyperbolic tangent on the right-hand side, instead has slope $4{\cal J}$ at the origin and approaches $\pm 1$ at infinity. The important point is that the values of the left-hand side function at $m_s = \pm 1$ are $\pm 2/(2-m^2)$ which lie above (respectively below for negative $m_s$) the hyperbolic tangent asymptote. This means that, for small beta satisfying $4 {\cal J} < 2/(1-m^2)$ the only solution to the first equation is $m_s=0$. For ${\cal J} > 1/[2(1-m^2)]$, instead, two new solutions $m_s\ne 0$ appear. As always, one expects the solution $m_s=0$ to become unstable at this point. 

To check this point let me compute the second-order partial derivatives of the free energy,
\begin{eqnarray}
{\partial^2 f_{MF} \over \partial m^2} &&= -4(2d-1) 
- 4{\cal J} (2d-1)^2 \left\{ 1- {\rm th}^2 (-) \right\} 
\nonumber \\
&&- 4{\cal J} (2d-1)^2 \left\{ 1- {\rm th}^2 (+) \right\} \ ,
\nonumber \\
{\partial^2 f_{MF} \over \partial m_s^2} &&= 4 - 4{\cal J} \left\{ 1- {\rm th}^2 (-) \right\} 
- 4{\cal J} \left\{ 1- {\rm th}^2 (+) \right\} \ ,
\nonumber \\
{\partial ^2 f_{MF} \over \partial m\partial m_s} &&= 4{\cal J} (2d-1) \left\{ 1- {\rm th}^2(-) \right\} 
\nonumber \\
&&- 4{\cal J} (2d-1) \left\{ 1- {\rm th}^2 (+) \right\} \ .
\label{thirteen}
\end{eqnarray}
where I have introduced the short-hand notation
\begin{eqnarray}
{\rm th}(+) &&= {\rm th} \left[ 2{\cal J} \left( (2d-1)m+m_s\right) -H\right] \ ,
\nonumber \\
{\rm th}(-) &&= {\rm th} \left[ 2{\cal J} \left( (2d-1)m-m_s\right) -H\right] \ .
\label{fourteen}
\end{eqnarray}

Since $\partial^2 f_{MF} /\partial m^2 < 0$, the sign of the determinant $(\partial^2 f_{MF} /\partial m^2) (\partial^2 f_{MF} /\partial m_s^2 ) - (\partial ^2 f_{MF} /\partial m\partial m_s)^2$ is determined entirely by 
$\partial^2 f_{MF} /\partial m_s^2$. Using (\ref{ten}) we can rewrite this as
\begin{equation}
{\partial^2 f_{MF} \over \partial m_s^2} \left( m_s=0\right) = 4-8{\cal J} \left( 1-m^2 \right) \ .
\label{fifteen}
\end{equation}
This shows that on the line ${\cal J} = 1/[2(1-m^2)]$, where the two new solutions $m_s\ne 0$ appear, the term 
$\partial^2 f_{MF} /\partial m_s^2$ changes sign from positive to negative when ${\cal J} $ is increased. Since, moreover $(\partial ^2 f_{MF} /\partial m\partial m_s) \left( m_s=0 \right) =0$, it is the whole determinant $(\partial^2 f_{MF} /\partial m^2) (\partial^2 f_{MF} /\partial m_s^2 ) - (\partial ^2 f_{MF} /\partial m\partial m_s)^2$ that changes sign from negative to positive on the line ${\cal J} = 1/[2(1-m^2)]$ when ${\cal J}$ is increased. Finally, since both  $(\partial^2 f_{MF} /\partial m^2) \left( m_s =0 \right) < 0$ and $(\partial^2 f_{MF} /\partial m_2^2) \left( m_s =0 \right) < 0$ for ${\cal J} > 1/[2(1-m^2)]$ all this shows that, indeed, the unique high-temperature solution $m_s=0$ becomes a local maximum when the two new solutions $m_s \ne 0$ appear: these are thus the new stable solutions and at the critical temperature ${\cal J}_c = 1/[2(1-m^2)]$ there is a phase transition from a high-temperature phase characterised by $m \ne 0$ and $m_s=0$ to a low-temperature phase with $m\ne 0$ and $m_s \ne 0$. 

We have already seen (eq. (\ref{fifteen})) that the second derivative of the free energy vanishes at the transition. Let me further compute the third- and fourth-order derivatives:
\begin{eqnarray}
&&{\partial^3 f_{MF} \over \partial m_s^3} = 16{\cal J}^2 \left[ {\rm th}(+) \left( 1- {\rm th}^2(+)\right) -{\rm th}(-) \left( 1- {\rm th}^2(-)\right) \right] \ ,
\nonumber \\
&&{\partial^4 f_{MF} \over \partial m_s^4} = 32 {\cal J}^3 \left[ \left( 1- {\rm th}^2(+) \right)^2 + \left( 1- {\rm th}^2(-) \right)^2 \right] 
\nonumber \\
&&+ 64 {\cal J}^3 \left[ {\rm th}^2(-) \left( 1- {\rm th}^2(-)\right) -{\rm th}^2(+) \left( 1- {\rm th}^2(+)\right) \right] \ .
\label{sixteen}
\end{eqnarray}
Using again eq. (\ref{ten}) one can easily conclude that 
\begin{eqnarray}
{\partial^3 f_{MF} \over \partial m_s^3} \left( m_s=0\right) &&= 0 \ ,
\nonumber \\
{\partial^4 f_{MF} \over \partial m_s^4} \left( m_s=0\right) &&= 32 {\cal J}^3 \left( 1-m^2 \right)^2 > 0 \ ,
\label{seventeen}
\end{eqnarray}
which shows that, for finite ${\cal J}$ (for which also $m<1$) the phase transition is of second-order, i.e. continuous, although one cannot exclude that this is an artefact of the mean field approximation. 

The phase boundary can be derived easily by using eqs. (\ref{ten}) for $m_s=0$ together with ${\cal J}_c = 1/[2(1-m^2)]$ and the inverse of the hyperbolic tangent: ${\rm th}^{-1}(x) = (1/2) {\rm ln} \left( (1+1)/(1-x) \right)$, 
\begin{equation}
H = (4d-2) {\cal J} m + {1\over 2} {\rm ln} {1+m\over 1-m} \ , \qquad  m=\sqrt{1-{1\over 2{\cal J}}} \ .
\label{eighteen}
\end{equation}
or, reintroducing the coupling constant $J$ of the original model,
\begin{equation}
h ={2d-1 \over 4}Jm+ {T\over 2} {\rm ln} {1+m\over 1-m} \ , \quad  m=\sqrt{1-{4T\over J}} \ .
\label{nineteen} 
\end{equation}

The phase boundary (\ref{nineteen}) in the $h$-$T$ plane, together with the magnetic field value implied by eq. (\ref{three}), are shown in Fig. 1 for $d=4$ and $J=1/10$.

\begin{figure}
\includegraphics[width=8cm]{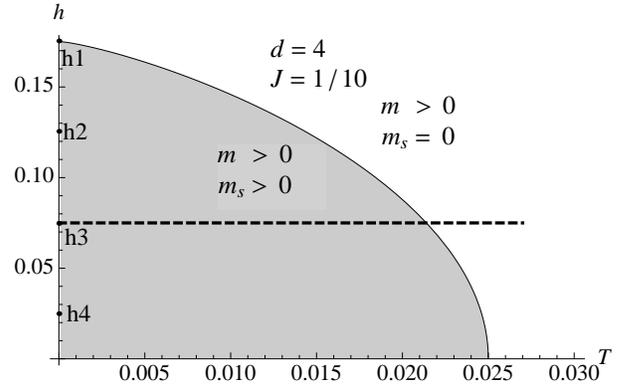}
\caption{\label{fig:Fig1} The phase boundary of the antiferromagnetic transition in the $h$-$T$ plane, together with the fixed magnetic field of the model, eq. (\ref{three}), for $d=4$ and $J=1/10$. The values $h_i$ ($i=1\dots 4$) represent the critical magnetic fields where $i$ edges decouple for each vertex at $T=0$.}
\end{figure}

At zero temperature and above the critical magnetic field $h_c = (2d-1)J/4$ (= 0.175 for $d=4$ and $J=1/10$), the mean field solution implies $m=1$ and $m_s=0$. As already stressed above, this means that the ground-state configuration is the entire graph on which the model is defined. To make contact with geometry, let me consider the simple example of a $d$-dimensional hypercubic lattice. In this case, the spectral dimension $d_s$ coincides with the Hausdorff dimension $d_H$, both being determined by the graph connectivity, $d_s = d_H = d$.

When the magnetic field is smaller than a second critical value $h_4= J/4$, instead, the mean field equations (\ref{ten}) imply a solution $m=0$, $m_s=1$. In the language of dynamical graphs this indicates a ground state consisting of a $d$-regular sub-graph that touches every vertex of the original $2d$-regular graph on which the model is defined. This can be viewed either as a higher-dimensional generalization of a Hamiltonian cycle \cite{ham} or as a self-avoiding hypersurface on the graph \cite{saw}. In the simple example of a hypercubic lattice, the spectral dimension is now clearly decreased to $d/2$, since a random walker has maximally $d$ edges to choose from at every step, $d_s=d/2$. The Hausdorff dimension of this dynamical sub-graph, instead remains $d_H=d$ since the sub-graph visits every vertex of the original graph (exactly once), i.e. it is space-filling in the original graph. 

Between the two critical values $h_4=J/4$ and $h_c=(2d-1)J/4$ the mean field equations (\ref{ten}) have no solution at $T=0$. This indicates the failure, in this region, of the Ansatz of a configuration with half the edges of one type and half of the other type.  It is easy, though, to repeat the mean field computation with a generic Ansatz in which, at every vertex, there are $n_d$ dark-coloured edges and $n_l$ light-coloured edges, with $n_d + n_l =2d$. The mean field equations become
\begin{eqnarray}
&&m+m_s = {\rm th} \left[ H-2{\cal J} \left( (2d-1) m - (1-(n_d-n_l)) m_s\right) \right] \ ,
\nonumber \\
&&m-m_s = {\rm th} \left[ H-2{\cal J} \left( (2d-1) m + (1+(n_d-n_l)) m_s\right) \right] \ . 
\label{twenty}
\end{eqnarray}
While the upper critical magnetic field $h_c=(2d-1)J/4$ remains unchanged, the lower value is now increased to 
$h_i = (1+(2d-2i)) J/4$ for $n_d=2d-i$ and $n_l=i$, $i=1\dots d$. This coincides with the upper value $(2d-1)J/4$ for $n_l=i=1$, $n_d=2d-1$ : $h_1=h_c$. This shows that, at $T=0$, the quantum phase transition is of first-order, implying a jump from $m=1$, $m_s=0$ to $m=0$, $m_s=1$ and the "decoupling of one edge", since $n_l=1$ and $m=0$, $m_s=1$ together imply that at each vertex one edge of a given (light) colour is absent while all others are present. In the hypercubic lattice example, this defines a regular sub-graph of spectral dimension $d_s= d-1/2$ and Hausdorff dimension $d_H=d$. It is easy to show that, lowering the magnetic field below the further critical value $h_2 = (2d-3) J/4$ at $T=0$, the free energy is minimized by the Ansatz $n_d=2d-2$, $n_l=2$. At this value an entire dimension (two edges) decouples. When $h$ is further decreased, at $T=0$, more dimensions decouple at $h_i$, $i=3\dots d$, until, at $h=J/4$ the spectral dimension hits its lowest value $d/2$, as described above. It is to be expected that these first-order quantum phase transitions extend as phase boundaries in the $T>0$ region all the way to the $m_s \to 0$ phase line. Unfortunately, the analysis of these phase boundaries for $T>0$ is much more complex and goes beyond the scope of the present paper. 
The important point here is, rather, that this toy model clearly shows how the statistical antiferromagnetic order parameter $m_s$ is related to the geometric disorder parameter $(d_H-d_s)$: when $m_s$ vanishes $d_H-d_s$ also vanishes, conversely, a non-zero value of $m_s$ indicates a decoupling between $d_s$ and $d_H$. 

For the values of $d$ and $J$ shown in Fig.1, the transition happens at finite temperature, since the magnetic field implied by the original model (\ref{two}) is a not a free variable but it is, rather, fixed itself in terms of the coupling constant $J$ as in (\ref{three}). This value is shown as an horizontal line in Fig. 1. The geometric transition described by the antiferromagnetic order parameter $m_s$ is thus a second-order (in mean field theory) transition at a critical finite temperature $T_c$. At this temperature the dynamical graph becomes critical, with long-range correlations between edges on the distance measure defined by the embedding graph distance. The idea of the original model \cite{trug} is that such a critical point defines a continuum space-time, exactly as in the original Kazakov model \cite{kazakov} but not anymore restricted to $2D$. 

Inserting the magnetic field (\ref{three}) into eq. (\ref{nineteen}) one can derive the phase boundary in terms of the original coupling constant $J$,
\begin{eqnarray}
J &&= {1\over (2d-1)(1+m)} \left( 1 - 2\ T\ {\rm ln} {1+m\over 1-m} \right) \ ,
\nonumber \\
m &&= \sqrt{1-{4T\over J}} \ .
\label{twentyone}
\end{eqnarray}
This shows that, for $J=1/(4d-2)$, the critical temperature becomes $T=0$ and the geometric critical point corresponds to a purely quantum phase transition, as is shown in Fig. 2. This is exactly as in the full model \cite{trug} with the only difference that there $J=1/(4d-1)$ and the quantum phase transition corresponds to $d=4$. In the present toy model this quantum phase transition is of first order, as explained above. Even a small temperature, however, is sufficient to smoothen out the transition and make it of second-order (in the mean field approximation). 

\begin{figure}
\includegraphics[width=8cm]{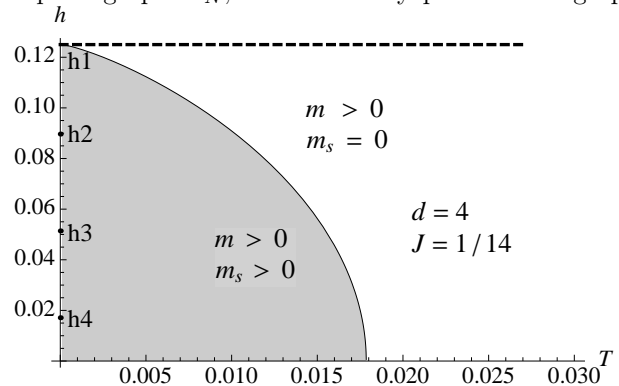}
\caption{\label{fig:Fig1} The phase boundary of the antiferromagnetic transition in the $h$-$T$ plane, together with the fixed magnetic field of the model, eq. (\ref{three}), for $d=4$ and $J=1/(4d-2)=1/14$. For this value of the coupling constant the geometric critical point corresponds to a purely quantum phase transition.}
\end{figure}

Of course, the original model \cite{trug} is defined on the complete graph $K_N$, to allow every possible sub-graph as a dynamical "space-time graph" and, unfortunately, the complete graph is not bipartite. As in usual antiferromagnets, the frustration relations on non-bipartite graphs are typically much subtler than on bipartite graphs and lead to much more complex behaviour. It is possible that the recently developed network tensor methods \cite{tensor} may shed some light on the critical behaviour of the fully frustrated model. For the moment, though, only numerical evidence is available in this case \cite{trug}. This numerical evidence, however, supported by recent analytical results on finite-size scaling in the Ising model \cite{berche} points indeed to the existence of a geometric quantum phase transition at $d_H=d_s=4$.

\end{document}